\documentclass[11pt]{article}  

\usepackage{amsmath,amsthm,latexsym,amssymb,amsfonts,epsfig}

\newcommand{\be}{\begin{equation}}
\newcommand{\ee}{\end{equation}}
\newcommand{\ba}{\begin{eqnarray}}
\newcommand{\ea}{\end{eqnarray}}

\title{{\sf Reduced phase space induced decay conditions}}
\author{
{\sf T. Thiemann}$^1$\thanks{{\sf 
thomas.thiemann@gravity.fau.de}}\\
\\
{\sf $^1$ Inst. for Quantum Gravity, FAU Erlangen -- N\"urnberg,}\\
{\sf Staudtstr. 7, 91058 Erlangen, Germany}\\
}
\date{{\small\sf \today}}

\makeatletter
\@addtoreset{equation}{section}
\makeatother

\begin{document} 

\maketitle

{\sf

\begin{abstract}
The definition of the phase space of field theories in presence of boundaries of Cauchy surfaces
requires a choice of boundary conditions or decay behaviour of those fields. Often these 
conditions are motivated in part by the decay behaviour of the initial data of known exact solutions. 

In the case of gauge field theories the initial data are not free but are subject to initial
value constraints. Still, the decay behaviour is commonly specified for the kinematical, 
i.e. unconstrained phase space. This can lead to the following practical
problem: The constraints are preferably solved for field variables on which they depend 
only algebraically, i.e. not involving derivatives, as otherwise one would need to solve 
partial differential equations. However, the specified decay behaviour may prevent from 
doing that.  

On the other hand, a precise specification of decay for all kinematical fields appears 
unnecessary because the decay of gauge degrees of freedom is not observable. Yet, knowledge 
of their decay is required as one needs to compute Poisson brackets on the kinematical 
phase space in order to define what gauge invariance means. Thus the interplay between 
the constraint structure and the decay properties of the kinematical phase space is complex.  

In this contribution we develop a reduced phase space induced approach to the decay 
problem. It requires as an input a choice of gauge conditions which is tailored to
the algebraic structure of the constraints. Together with the constraints,
these define a split of the kinematical phase space into gauge and true degrees of freedom.
Then the decay conditions of the kinematical phase space is systematically parametrised
by a choice of decay for just the true degrees of freedom (i.e. the reduced phase space),
the decay of the gauge degrees of freedom then follows unambiguously from solving 
both the constraints and the gauge conditions.
\end{abstract}

\section{Introduction}
\label{s1}

Field theories defined on a spacetime manifold with boundaries require a specification of 
boundary conditions. For example, in order to derive the Euler-Lagrange equations from
an action principle one requires that the field variations decay sufficiently fast 
at both spatial and temporal boundaries of the spacetime manifold in order to carry 
out the integrations by parts. Similarly, in the Hamiltonian approach one 
specifies the decay behaviour of the fields and their conjugate momenta at spatial
boundaries of an initial data (Cauchy) surface. There is considerable freedom in the 
choice of such decay conditions as one only needs to require that the symplectic (or 
Poisson bracket) structure 
and the Hamiltonian have 1. convergent integrals over the Cauchy surface and 2. well-defined
functional derivatives. The second requirement sometimes requires to add a boundary (counter)
term to the bulk Hamiltonian. The choice of decay is not uniquely defined by these 
requirements and requires additional physical input, often guided by asking that the 
solutions of the resulting (Hamiltonian) equations of motion with the specified class of 
initial data include known solutions of the Euler-Lagrange equations.\\
\\
The above description is complete for field theories without gauge symmetries. Gauge field theories 
are in addition subject to constraints on initial data in their Hamiltonian description. 
In order to keep the presentation concise, we will assume that the theory is totally constrained, i.e. there is no a priori 
distinguished Hamiltonian and all constraints are of first class (one can always pass to that situation by 
enlarging the phase space to absorb a possibly present Hamiltonian into an additional constraint 
and by passing from the Poisson to the Dirac bracket to remove second class constraints).
This has several non-trivial consequences for the resulting phase space description (more details 
can be found in e.g. \cite{1}):\\
i. \\
By definition, in a gauge theory not all degrees of freedom are observable. Therefore 
imposing decay behaviour on all (kinematical) degrees of freedom appears to be unnecessary
as the decay of non-observable gauge degrees of freedom is anyway not detectable. 
On the other hand, without specifying the decay also on the gauge degrees of freedom 
one cannot define Poisson brackets on the kinematical phase space. Since an observable 
is a function on the kinematical phase space which has (weakly) vanishing Poisson brackets 
with the constraints (which are also functions on the kinematical phase space), the very 
definition of what is an observable requires access to the decay behaviour of all fields.
Therefore the best one can hope for is that the definition of an observable is invariant 
under a change of decay behaviour of the gauge degrees of freedom.\\
ii.\\
Part of solving the (Hamiltonian) equations of motion consists in solving the constraints.
These typically depend algebraically on the momentum degrees of freedom or at most on their 
first partial derivatives while they depend on second partial derivatives of the configuration 
degrees of freedom. It is therefore much simpler to solve the constraints for the momenta.
Now suppose that the specified decay behaviour on the kinematical phase space 
is such that the momentum degrees of freedom contribution to the constraints decays faster than 
than the configuration degrees of freedom contribution as we approach the boundary. Then one cannot 
solve the constraints for the momenta! In this case one would then need to solve a second 
order partial differential equation for configuration degrees of freedom
rather than an algebraic or first order one. While 
not a problem in principle, it is a problem in practice as sufficiently concrete formulae 
are necessary to have at one's disposal for further developments of the theory such as 
solving the equations of motion, deriving a gauge invariant Hamiltonian or quantisation.\\  
iii.\\
The constraints are smeared with test functions. If these are of compact support away from 
the boundaries, these smeared constraints are functionally differentiable and generate gauge transformations
via Poisson brackets. However, one is also interested in more general test functions not 
of compact support away from the boundaries. When the constraints are smeared with these 
more general test functions, they typically fail to be functionally differentiable and have to be 
augmented by counter boundary terms. Whether this is possible depends critically on the decay 
behaviour of all fields (initial data and smearing functions). When possible, technically 
the deviation of the variation of the constraints from a bulk term has to be the variation of a 
boundary term, these 
augmented functionals are no longer constrained to vanish but have (weakly) vanishing Poisson brackets 
with the constraints smeared with test functions of compact support away from the boundaries,
due to the first class property of the constraints. Hence these augmented constraints
define observables and generate symmetry transformations on all observables rather than gauge transformations.
Since typically 
the smearing functions also have a spacetime interpretation, one often chooses the decay of those 
more general smearing functions tailored to known solutions of the Euler-Lagrange equations. 
It is a priori however unclear whether any of those symmetry generators plays the role of the 
physical Hamiltonian generating the equations of motion of the observable degrees of freedom.   
A systematic approach would be desirable that derives the decay behaviour of those smearing fields
corresponding to the physical Hamiltonian.\\
\\
It transpires that there is a complex interplay between the choice of decay on initial data 
and the constraints in a gauge field theory. On the one hand, practical considerations 
forbid a completely free assignment of decay on the full kinematical phase space, on the other hand 
the observable physics should be insensitive to the details of the decay of the gauge sector of the theory.
Obviously, the key to a systematic approach is a clean separation between the treatment 
of the observable and gauge sector of the theory. By definition, a (so called Dirac) observable has 
(weakly) vanishing Poisson brackets (computed on the kinematical phase space) with 
the constraints smeared with test functions of compact support (or rapid decay) away 
from the boundaries. 

This definition is simple but it is not practically useful 
because the corresponding Poisson bracket equations are too complicated to solve for 
field theories of interest. 
A (locally in phase space) equivalent approach uses as an input a choice of gauge fixing 
conditions. There as many gauge fixing conditions as there are constraints and the 
practical considerations made above motivate to choose them as conditions on those configuration degrees
of freedom which are conjugate to those momenta that one wants to solve the constraints for. 
We call these canonical pairs singled out by gauge fixing and solving the constraints the 
``gauge'' sector, the remaining canonical pairs the ``true'' sector. Of course, these 
true variables do not have weakly vanishing Poisson brackets with the constraints,
hence they are not Dirac observables. However, as one can show \cite{1,2}, there exists 
a unique Dirac observable corresponding to each of these true variables
which is constructed by a closed formula using both gauge fixing condition and constraints.
Moreover, the Poisson brackets between these so called relational observables \cite{3}
are weakly isomorphic to the Poisson brackets of the corresponding true variables. 
Clearly the whole construction requires as an additional input a choice of gauge fixing 
condition, however, it enters simply as a practical tool to construct Dirac observables
and by no means indicates a gauge dependence of the description. Using different 
gauge fixing conditions means constructing different relational Dirac observables but 
for any such choice of gauge fixing conditions the totality of Dirac observables 
is a complete description of the observable sector of the theory. Since the true 
and (relational) observable phase space are isomorphic, we may as well pick the much 
simpler description in terms of gauge and true sector specified by a gauge fixing 
condition. We therefore consider ``true'' and ``observable'' as synonymous for what 
follows.\\
\\     
With these considerations out of the way, the idea is to use this ``reduced phase space'' point 
of view to approach the specification of decay on the kinematical phase space. 
It requires as an input 1. a choice of gauge fixing conditions tailored to the algebraic properties 
of the constraints as sketched above which allows the split into the gauge and observable sector and 2.
a choice of decay just for the observable sector. After this, everything else follows:\\
1.\\
We evaluate the constraints at the chosen gauge fixing condition and solve the constraints 
for singled out gauge momenta which yields functions 
depending on the true variables whose decay can therefore be evaluated. We define the decay of 
the gauge momenta to coincide with the decay of those functions.\\
2.\\
We define the decay of the gauge configuration variables to be such that the 
gauge fixing condition decays rapidly at the boundaries. Together with 1. this implies 
that the entire kinematical symplectic structure is well defined.\\
3.\\
A key role is played by the solutions of the stability conditions. These are those smearing 
functions of the constraints which have the property that 
the constraints smeared with them have vanishing Poisson 
brackets with the gauge fixing conditions when both constraints and gauge fixing conditions hold.
In general these correspond to symmetry transformations and thus require the afore mentioned
counter terms to the constraints. At this point, one must self-consistently tune 
the decay behaviour of the observable sector such that the boundary term picked
up in a variation of the constraints is a variation even when the smearing function
has the same decay as a solution of the stability condition. \\
4.\\
After this is achieved, the physical Hamiltonian precisely corresponds to that symmetry 
transformation which is generated by a solution of the stability condition.\\ 
\\
\\ 
This contribution is organised as follows:\\
\\
\\

In section \ref{s2} we begin motivating the general considerations of the present paper with 
the example of the asymptotically flat sector of vacuum General Relativity and the usual 
kinematical boundary conditions employed in that context. 

In section \ref{s3} we go in more details and state the problem sketched in the introduction 
using more mathematical structure. 

In section \ref{s4} we use that technology to formulate a self-consistent concrete algorithm 
in order to solve the problem stated in section \ref{s3}..

In section \ref{s5} we summarise and conclude.

\section{Motivating example}
\label{s2}

The considerations of the present work are motivated by General Relativity (GR) in 
the asymptotically flat sector. We will only consider vacuum GR or equivalently assume that 
the decay of the matter fields is faster than those of geometry. It will be 
sufficient to consider the case of 3+1 spacetime dimensions, the generalisation to other dimensions 
being straightforward. Of course, the spacetime is supposed to be globally hyperbolic 
as otherwise the phase space formulation is not available.\\

In this situation the phase space 
is encoded by a 3-metric $q$ and its conjugate 
momentum $p$ on Cauchy slices $\sigma$ ($p$ is related to the extrinsic curvature 
of $\sigma$) and chooses boundary conditions at spatial infinity 
which encompass the condition that $q$ becomes the flat Euclidian metric $q^0$. 
Using the asymptotic Cartesian coordinates defined by $q^0$ with respect to which
it becomes the Kronecker symbol, one then imposes conditions on the kinematical 
phase space coordinates $q-q^0,\; p$ as $r\to\infty$ where $r$ is the Cartesian radial
coordinate. A popular choice \cite{4} is that $q-q^0$ is an even parity tensor on the asymptotic 
sphere $\partial\sigma$ which decays as $1/r$ while $p$ is of odd parity 
and decays as $1/r^2$ (both can have corrections of higher order in $1/r$ 
of any parity). Here parity is with respect to the standard reflection on the sphere.
This is motivated by the Schwarzschild solution which falls in this 
decay class (in that case $p=0$ as 
the metric is static in these coordinates). 

As outlined in the introduction, we note that here one imposed a decay 
behaviour on the entire kinematical phase space without paying attention to the constraint 
structure. Now the initial value constraints of vacuum GR are the vector 
constraint $V=-2\nabla\cdot P=0$ and the Hamiltonian constraint 
$K=P\cdot G(q)\cdot P-\det(q) R(q)=0$ where $\nabla q=0$ is the torsion free 
covariant differential compatible with $q$, $G(q)$ is the DeWitt metric (a tensor 
of fourth rank bilinear in $q$) and $R(q)$ is the Ricci scalar of $q$, see \cite{5} 
for details.
Plugging in the decay behaviour just stated, one finds that $R(q)=O(1/r^3)$ while 
$\det(q),\; G(q)=O(r^0)$ and $P^2=O(1/r^4)$. This means that one cannot simply solve 
$K=0$ for $P$, rather the $O(1/r^3)$ terms must cancel separately. Thus without 
further input supplied by a gauge fixing condition in this case 
one would be forced to solve $K=0$ for certain components of $q$ rather than $P$
which in this case is rather difficult because $R(q)$ involves up to second partial derivatives 
of $q$ and thus amounts to solving difficult PDE's for $q$ rather than simple algebraic equations for $P$. 

In order to avoid this, one can instead follow the strategy sketched in the introduction 
and deviate from the 
boundary conditions of \cite{4} by specifying the decay only on a choice of true 
degrees of freedom $(Q,P)$ singled by a set of gauge fixing conditions on a set of configuration 
gauge degrees of freedom $x$ conjugate to gauge momenta $y$ where $(q,p)=((x,y),(Q,P))$ is 
the corresponding split of the canonical chart. Then an $O(1/r)$ respectively $O(1/r^2)$ decay of the 
true degrees of freedom $(Q,P)$ with above parity restrictions maybe still freely specified  
while $y$ may pick up a weaker decay depending on the chosen gauge condition on $x$ as dictated 
by the constraints. 

Only if one imposes a gauge fixing condition that amounts to saying that the 
$O(1/r^3)$ terms of $R(q)$ are not present can one one maintain the decay behaviour \cite{4}
on all $p$ while being able to solve $K=0$ for $y$. 
A possible such gauge fixing condition consists in the symmetric, trace free, transverse 
(STT) gauge \cite{6}. Here one splits $q=q^0+Q+x$ where $Q$ is symmetric, has vanishing trace 
Tr$([q^0]^{-1} Q)=0$ and vanishing divergence $[q^0]^{-1}\nabla_0 \cdot Q$ with respect 
to the flat Euclidian metric where $\nabla^0 q^0=0$. The tensor $x$ contains the 
longitudinal and trace information. The conjugate momenta $P,y$ have the same properties. 
In that case it is not difficult to see that we can have $Q=O(1/r)$ of even parity and $P=O(1/r^2)$ of odd parity such 
that $R(q)=O(1/r^4)$ when $x=0$ or decays rapidly as $r\to \infty$ and thus $K=0$ can be solved algebraically for $y$ while 
$y=O(1/r^2)$. In fact, it is not necessary to install the full STT gauge in order to achieve this,
a single gauge condition that relates the trace to the doubly longitudinal component of $q-q^0$ 
would be sufficient and both can still have $O(1/r)$ decay. 
We encounter such weaker versions of the STT conditions in black hole perturbation theory \cite{7}.

\section{The interplay between constraint and decay structure of a phase space} 
\label{s3}

Having motivated the research question of the present work via a concrete field theory in the previous section, we now investigate the 
interplay between constraints and decay in a general setting providing details left over in the introduction. The 
Cauchy surface is denoted by $\sigma$ and its boundary by $\partial\sigma$.\\
\\  
1.\\
Let $(q^a(z),p_a(z))$ denote the points of the unconstrained or kinematical phase space where $q^a$ are 
the spacetime fields pulled back to the Cauchy surface and $p_a$ their conjugate 
momenta. Here $a$ labels discrete and finite information (e.g. tensorial indices) while $z$ 
denotes points in the initial data surface $\sigma$. A minimal requirement  
on the decay behaviour of the fields at $\partial\sigma$ is that the integral 
defining the symplectic potential $\Theta=\int_\sigma\; dz\;p_a(z)\;[\delta q^a](z)$
converges.

The presence of initial data constraints $C_I(q,p;z)=0$, 
where $I$ takes values in another finite and discrete label set, implies 
that the data $(q^a(z),p_a(z))$ are not entirely free, i.e. one cannot independently pose 
initial conditions on all degrees of freedom of the kinematical phase space. We will only 
consider the case of first class constraints, i.e. their Poisson brackets on the kinematical 
phase space are linear 
combinations of constraints (with coefficients that possibly depend on the phase space 
point). We also only consider the case of totally constrained systems without Hamiltonian
(as already mentioned, one can always achieve this by extending the kinematical phase space by one canonical 
pair $(q^0,p_0)$ and adding the constraint $C_0=p_0+h$ if there is an a priori Hamiltonian 
density $h\not=0$).  \\
2.\\
The decay conditions on $q^a(z),p_a(z)$ at $\partial\sigma$ have to be augmented 
by the requirement that the smeared constraints $C(S)=\int_\sigma\;dz\; S^I(z)\; C_I(z)$
are convergent and functionally differentiable where $S^I$ are test functions not depending 
on $(q,p)$ For $S^I$
of compact support away from $\partial\sigma$ this requirement is easily met. However,
for more general $S^I$ this typically requires to add a boundary term $B(S)$ to $C(S)$ 
resulting in $H(S)=C(S)+B(S)$. The boundary term arises as follows: One allows a decay of $S$
such that the variation $\delta C(S)$
differs from a bulk integral of the form 
$\int_\sigma\;dz\; [J^S_a(z)\; [\delta q^a](z) + J_S^a(z)\; [\delta p_a](z)]$ 
by a variation 
$-\delta B(S)$ where $B(S)$ is an integral over $\partial\sigma$. One calls canonical transformations generated by 
$H(S)$ for which $B(S)=0$ a gauge transformation,
otherwise a symmetry transformation. The above closure condition then means that 
$\{H(S),H(S')\}=H(\kappa(S,S'))$ for certain structure functions $\kappa:\;{\cal S}_0\times {\cal S}_0
\to {\cal S}_0$ where ${\cal S}_0$ corresponds to the space of those $S$ for which $H(S)$ generates
a gauge transformation and $\{.,.\}$ denotes 
the Poisson bracket on the kinematical phase space. The function $\kappa$ typically extends 
to the larger space $\cal S$ with values in $\cal S$ corresponding to those $S$ for which 
$H(S)$ generates a symmetry transformation. \\ 
3.\\
Not all kinematical degrees of freedom carry observable information, the kinematical 
phase space description is redundant. This means that one can impose as many {\it gauge conditions} 
$G^I(q,p;z)=0$ on the kinematical phase space as there are constraints. An 
admissible gauge condition has two properties: 1. At least locally in the kinematical 
phase space, it is possible 
to split the points of the kinematical phase space into pairs $(q^a,p_a)=((x^I,y_I),\;(Q^A,P_A))$ 
such that one can solve the sets of equations $C_I(.,.;x)=G^I(.,.;x)=0$ uniquely for 
$x^I(z)=X^I(Q,P;z),\; y_I(z)=Y_I(Q,P;z)$. 2. Given a point $(q^a(z),p_a(z))$ on the constraint
surface $C_I(z)=0$ in the vicinity of $((X_I(Q,P;z),Y^I(Q,P;z)),(Q^A(z),P_A(z)))$ it is possible 
to find unique smearing functions $S^I(z)$ corresponding to a gauge transformation 
such that $\{H(S),G^I(z)\}=-G^I(z)$. The uniqueness condition is satisfied when 
the {\it Dirac kernel} $\Delta_I\;^J(z,z'):=\{C_I(z),G^J(z')\}$ is non-degenerate.  
We call $(x,y)$ the {\it gauge} degrees of freedom and 
$(Q,P)$ the {\it observables} or {\it true} degrees of freedom. The latter are canonical
coordinates of the
{\it reduced phase space}. Note that the notion of gauge and true
depends on the choice of gauge condition. As mentioned in the introduction,
one can show that nevertheless there is an isomorphic
description in terms of Dirac observables and different gauge choices simply correspond 
to different sets of Dirac observables which capture the same observable information in total.\\
4.\\
By the uniqueness requirement of 3., there is no non-trivial gauge transformation satisfying $\{H(S),G^I(z)\}=0$ when 
$G^I(z)=C_I(z)=0$. However, there may be a symmetry transformation that does. We denote  
by $S^I_\ast(Q,P;z)$ the solutions of the {\it stability condition} $\{H(S),G^I(z)\}_{C=G=0}=0$. 
Typically the $S^I_\ast$ are elements of a finite dimensional linear space parametrised 
by a finite number of parameters corresponding to integration constants of the partial 
differential equations that have to be solved.
The {\it physical or reduced Hamiltonian}, if it exists, corresponding to a solution $S^I_\ast$ is 
a function $E(Q,P)$ of the true degrees of freedom such that for any function $F(Q,P)$ 
of the true degrees of freedom we have $\{E,F\}=\{H(S),F\}_{S-S_\ast=C=G=0}$. A sufficient 
condition for existence is specified in \cite{8}: $B(S)$ can always be written as 
$B(S)=\int_{\partial\sigma}\;d\Omega \; S^I\; j_I$ where $\Omega$ is a measure on $\partial\sigma$ 
and $j_I$ are certain phase space dependent functions on the boundary. Suppose that 
$[S^I_\ast]_{|\partial\sigma}=[\frac{\delta \chi[j]}{\delta\; j_I}]_{j=j^\ast}$ where 
$j_I^\ast=[j_I]_{G=C=0}$, the functional
derivative is with respect to $\partial\sigma$ rather than the bulk $\sigma$ and $\chi$ is a function 
just of the $j_I$. Then $E=\chi[j^\ast]$ when $G^I$ take the form of coordinate 
gauge conditions (see item B. below).\\
\\
We realise that in the case of gauge theories with boundaries there is a complicated interplay between 
the choice of decay behaviour on the kinematical phase space and the very definition of the 
dynamics. In particular, we observe the following:\\
A.\\
On the one hand, one must be careful that the choice of decay conditions is not in conflict
with the constraints in the following sense: At least locally we can always solve the 
constraints $C_I=0$ for $y_I(z)=-h_I(x,Q,P;z)$. Now plugging in the chosen asymptotic behaviour 
of $x,Q,P$ as $z\to\partial\sigma$ into $h_I(x,Q,P;z)$ may result in an asymptotic decay of 
$-h_I$ different from the one chosen for $y_I$. If one maintains both the chosen decay behaviour
and the constraints one concludes that one cannot solve $C_0=0$ for $y_I=-h_I$ near $\partial\sigma$,
rather one needs to solve $C_I=0$ for variables different from $y_I$. Vice versa, if one wants to 
solve the constraints for $y_I$ as it maybe useful for practical reasons, one must either adapt 
the chosen decay behaviour to that purpose or one must achieve that the particular function 
$h_I$ decays in accordance with that of $y_I$. The latter maybe achieved using adapted gauge conditions. \\  
B.\\
On the other hand, imposing a specific decay behaviour on all kinematical degrees of freedom is 
also redundant in the following sense: Suppose we pick gauge conditions of the 
form $G^I(z)=x^I(z)-k^I(z)$ 
where $k^I(z)$ are certain functions on $\sigma$ which do not depend on the kinematical 
phase space point (such gauge conditions are therefore referred to as coordinate gauge conditions).
Then $(G^I,y_I)$ is a conjugate pair and we may impose arbitrarily strong decay on $G^I$ while 
allowing arbitrarily weak decay of $y_I$ at $\partial\sigma$. This is because e.g. rapid decay 
of $G^I$ is trivially consistent with the gauge condition $G^I=0$ and then the contribution 
$\int_\sigma\;dz \; y_I(z)\;\delta x^I(z)=\int_\sigma\; dz \; y_I(z)\;\delta G^I(z)$ to the 
symplectic potential trivially converges. This means
that whatever the decay of $h_I$ near 
$\partial\sigma$, it will be in agreement with that of $y_I$ and thus one can indeed 
solve the constraints for $y_I$.\\
C.\\
We realise that both A. and B. work nicely together in the following sense: Pick any 
decay of $Q,P$ such that the contribution 
$\int_\sigma \; dz\; P_A(z)\;\delta Q^A(z)$ to the 
symplectic potential converges. Then in terms of the functions specified above 
we have $X^I=k^I$ and $Y_I=-h_I(x=k,Q,P)$. We now {\it derive} the decay of $y_I$ 
to be that of $Y_I$ and {\it define} the decay of $x^I-k^I$ to be arbitrary, it just 
needs to be sufficiently strong in order that $\int_\sigma\; dz \; y_I(z)\;\delta G^I(z)$
converges. This solves both problems A. and B. in one stroke and displays the 
decay behaviour of the fields of the kinematical phase space as being parametrised by 
just that 
of the reduced phase space while being consistent with the desire to solve the constraints
for $y-I$. \\
D.\\
Furthermore, if we specify as decay that 
$x^I(z)-k^I(z)$ be of compact support away from $\partial\sigma$ then we will 
find $S^I(z)$ of compact support away from $\partial\sigma$ such that $\{H(S),G^I(z)\}=-G^I(z)$
when $C_I(z)=0$ 
because the right hand side of this equation depends only on $S^I(z)$ and finitely many of its 
derivatives for any field theory of physical interest. As such $S^I$ correspond to gauge 
transformations, the gauge $G^I=0$ is indeed reachable. 

\section{A possible solution strategy}
\label{s4}

The remarks made in the previous section suggest the following strategy towards specifying the decay 
of phase space fields $(q,p)$ constrained by first class constraints $C$
on Cauchy surfaces $\sigma$ with boundaries $\partial\sigma$: 
\begin{itemize}
\item[I.] We split the kinematical phase space conjugate pairs $(q,p)$ into two sets
$(x,y)$ and $(Q,P)$ subject to the condition that the {\it Dirac kernel}          
$\{C,x\}$ is non-degenerate and that $C=0$ is efficiently (preferably algebraically) solvable for 
$y$. Since at this point we have not yet specified any 
boundary conditions we define the Dirac kernel by the weak derivative 
$\{C(S),x^J(z')\}:=\int\; dz\; S^I\;\{C_I(z),x^J(z')\}$ i.e. we use smooth $S^I$ decaying rapidly 
at $\partial\sigma$ so that we can ignore boundary terms when integrating by parts.
\item[II.] We impose a gauge fixing condition $G^I:=x^I-k^I$ where $x^I_\ast:=k^I$ are 
functions on $\sigma$ which do not depend on $q,p$.
\item[III.] By (some infinite dimensional version of) the inverse function theorem
the non-degeneracy of the Dirac kernel implies that we can uniquely solve $C_I=0$ 
for $y_I(z)=-h_I(x,Q,P;z)$ for each $(Q,P)$ at least sufficiently close to $x=x_\ast$.
In particular we are interested in $y_I^\ast(z):=-h_I(x_\ast,Q,P;z)$. Note that in contrast 
to $x^I_\ast$ the function $y_I^\ast$ necessarily depends non-trivially on $(Q,P)$.
\item[IV.] Ignoring any boundary contributions due to integrations by parts, 
we compute the solutions $S^I_\ast$ of the 
stability conditions $\{C(S),G^I\}_{G=C=0}=0$. These solutions may not decay rapidly at 
$\partial\sigma$ so that ignoring boundary contributions has to be justified as we will below
in step VII. Step IV. can therefore be carried out before subsequent steps because those 
steps in particular grant that $S^I_\ast$ is also rigorously obtained from 
$\{C(S)+B(S),G^I\}_{G=C=0}=0$, see below. 
\item[V.] At this point we pick decay conditions $D$ at $\partial\sigma$ on the $Q,P$ only.
A priori, these are subject only to the condition i. that $\int\; dz\; P_A(z)\;\delta Q^A(z)$ converges. 
We define the decay of $y_I$ to coincide with that of $y_I^\ast$ and we define the decay 
of $x^I-x-I^\ast$ to be rapid. In that way, the decay of all fields $(q,p)$ is parametrised 
by $D$ only (and of course by the split made in step I.).   
\item[VI.] Given $D$ we can now determine the decay $D'=D'(D)$ of $S^I_\ast$. We tune $D$ such 
that for phase space independent functions $S^I$ with decay behaviour $D'$ ii. the integral 
$C(S)$ converges for general $(q,p)$ on the full phase space with decay determined by $D$ 
specified in V., iii.
the variation $\delta C(S)$ can be written as a bulk integral involving 
$\delta q, \delta p$ (see item 2. of the previous section)
and a boundary integral term which however can be written as 
$-\delta B(S)$ and iv. also $B(S)$ converges on the full phase space. 
Then $H(S):=C(S)+B(S)$ is functionally differentiable even for $S$ with decay behaviour $D'$.
\item[VII.]
The solution $S^I_\ast$ obtained only formally in step IV. is now rigorously obtained from 
the condition\\ $\{H(S),G^I\}_{G=C=0}=0$. By construction, this formula is well defined for any $S$ of 
decay $D'$ and thus can be evaluated at $S=S_\ast$ which {\it is} of that decay by construction.  
\item[VIII.] If possible we further tune $D$ such that v. $[S^I_\ast]_{\partial\sigma}=
[\frac{\delta \chi}{j_I}]_{G=C=0}$ for some suitable $\chi$ where 
$B(S)=\int_{\partial\sigma} \; d\Omega \; S^I\; j_I$ (see item 4. in the previous section). 
Then $E=\chi[j^\ast],\;j_I^\ast=[j_I]_{G=C=0}$ is the physical Hamiltonian. 
\item[IX.]
If there is any freedom in $D$ left, pick it such that the set of solutions of the 
physical equations of motion obtained from $E$ or equivalently via 
$\dot{F}:=\{H(S),F\}_{S=S_\ast,G=C=0}$ and $C=0=G=0$ contain desired exact solutions 
of the Euler-Lagrange equations.
\end{itemize}
What makes this algorithm non-trivial is that $S^I_\ast$ has to be computed at first 
formally, i.e. prior to specifying the decay $D$ of the true degrees of freedom $(Q,P)$ 
and prior to having available a functionally differentiable version $H(S)$ of $C(S)$. 
Then $D$ is tuned self-consistently such that conditions I.-V. are met and the formal 
step to obtain $S^I_\ast$ is justified a posteriori. The algorithm offers a systematic 
and minimalistic approach towards choosing decay conditions on the kinematical phase space 
in the presence of Cauchy surfaces with boundaries and constraints guided by selection 
principles of outmost practical importance.

\section{Conclusion}
\label{s5}

In this short note we have tried to motivate a shift of perspective when imposing decay 
conditions on the field coordinates of a constrained phase space in the presence of boundaries
of initial data surfaces. While the usual approach is based on stating decay conditions 
irrespective of the concrete structure of the constraints, we have advertised a constraint
structure driven Ansatz which is motivated by practical considerations. 

These practical 
considerations start from the remark that the reduced phase space formulation of 
constrained field theories using gauge fixing conditions is the most economic platform 
for all kinds of applications, from the classical, numerical integration of the equations 
of motion to the canonical quantisation of the system. In the reduced phase space approach 
one solves the constraints classically and thus one is interested to render this step
as tractable as possible. This singles out certain degrees of freedom, typically momenta 
that we have denoted by $y$ in this work over other choices of variables that one might 
want to solve the constraints for. Then it is natural to impose gauge fixing conditions $G$ on 
the configuration variables $x$ conjugate to $y$. As $(x,y)$ are pure gauge degrees of freedom,
we do not care about the decay behaviour of those not observable variables in this approach.
We only care about the decay of the remaining, observable variables $(Q,P)$. Having 
specified a decay $D$ of $(Q,P)$ the decay of $(x,y)$ is in fact dictated by $D$ to be 
such that $y$ decays as the solution $y^\ast$ of the constraints $C=0$ when $G=0$ and 
that $x$ decays such that $G$ decays rapidly. Further considerations associated 
with the physical Hamiltonian $E$ defining the dynamics of functions $F$ of $Q,P$
via $\{E,F\}:=\{H(S),F\}_{S=S_\ast,C=G=0}$ where $S_\ast$ solves $\{H(S),G\}_{S=S_\ast,G=C=0}=0$
further confine the possible choices of $D$. Here $H(S)=C(S)+B(S)$ where $C(S)$ is the constraint 
smeared by $S$ and $B(S)$ is a boundary term which is self-consistently generated by asking 
that $H(S)$ is a functionally differentiable version of $C(S)$ even when $S$ decays as $S_\ast$
rather than rapidly at $\partial\sigma$. 
  
We believe that this new perspective has practical advantages over the more traditional     
approach towards stating the decay properties of the kinematical phase space and will 
explore its consequences on concrete future applications such as black hole perturbation theory
\cite{9}.

}

\end{document}